\documentclass[
 reprint,
 superscriptaddress,
 amsmath,amssymb,
 aps,
 prl,
]{revtex4-2}

\usepackage{graphicx}
\usepackage{bm}
\newcommand{\rf}{\mathring}

\begin{document}

\title{Pinching an elastic potato chip}

\author{Kosei Kataoka}
\affiliation{Department of Mechanical Engineering, Keio University, Yokohama, Kanagawa 223-8522, Japan}
\author{Tomohiko G. Sano}
\email{sano@mech.keio.ac.jp}

\affiliation{Department of Mechanical Engineering, Keio University, Yokohama, Kanagawa 223-8522, Japan}

\date{\today}

\begin{abstract}
Thin shells deform by transmitting stress across their surface, yet how a shell of general shape carries a localized load remains unclear. Combining experiments, simulations, and theory on pinched saddle-shaped elastic shells, we show that the asymptotic lines, curves of zero normal curvature, dictate both deformation and stiffness. Isometry turns the shell mid-surface into a medium obeying a wave equation in space. Material points lit by the pinch are lifted, while those in its shadow sink. Finite thickness blurs the lit--shadow boundary, and the same rule maps where a clamped shell is rigid. Our findings indicate that light and shadow offer a unified framework for understanding mechanical stress propagation along shell structures.
\end{abstract}

\maketitle

Propagation of light, information, and materials is a fundamental concept underlying nature~\cite{Whitham1974}.
Hyperbolic partial differential equations (PDEs), such as the wave equation, describe propagation in acoustics, electromagnetism, and fluid dynamics, where disturbances travel along characteristic lines~\cite{CourantHilbert1962}.
The stresses in a plastically deforming portion of a granular system obey hyperbolic PDEs~\cite{Geng2001}, and models of static packings likewise predict wave-like stress propagation along the characteristic lines (``light cones'')~\cite{Bouchaud1995, Wittmer1996, Cates1998}.
Elliptic PDEs, by contrast, have no characteristics. The two classes distinguish auxetic (conformal) from non-auxetic (wave-like) responses in mechanical metamaterials~\cite{Zheng2022, Czajkowski2022}. Wave transmission likewise reflects geometry, as in flexural waves in shells~\cite{EvansLevine2013} and surface Rayleigh waves in isostatic lattices~\cite{KaneLubensky2014}.

Mechanical responses of thin shells depend on the nature of their reference geometry; in particular, the sign of their Gauss curvatures, $K_{\rm G}$, alters the point-indentation response~\cite{Vaziri2008}. Circular or polygonal ridge formation occurs near the pole of dome-like shells $(K_{\rm G}>0)$~\cite{Pogorelov1988, Nasto2013, Gomez2016, Abe2026}, whereas stress propagates along the characteristics of cylindrical shells $(K_{\rm G}=0)$ or saddle-like shells $(K_{\rm G}<0)$~\cite{Darboux1896, Sabitov1992, AudolyPomeau2002}.
Given that stretching is costly for thin shells, isometric bending dominates and stretching localizes~\cite{AudolyPomeau2010}, giving rise to geometric-rigidity phenomena~\cite{Lazarus2012} such as ridges in crumpled paper~\cite{Lobkovsky1995, Witten2007}, wrinkles~\cite{CerdaMahadevan2003, Tobasco2022}, and snap-through~\cite{Pandey2014, Abe2025}.
Nevertheless, the fundamental mechanism of the shell stiffness across general shell structures, regardless of the sign of Gauss curvature, has been overlooked.

\begin{figure}[!b]
    \centering
    \includegraphics[width=\linewidth]{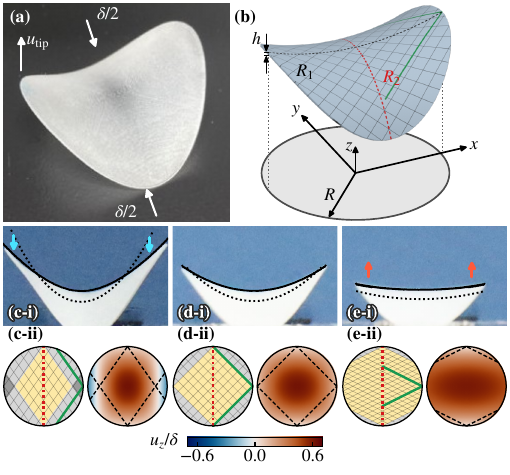}
    \caption{Problem definition.
    (a)~A saddle-shaped shell is pinched at two edge points by $\delta/2$ each, and its tip moves by $u_{\rm tip}$.
    (b)~Simulated shell of thickness $h$ and radius $R$ ($R_1/R_2=2$, $\delta=5$\,mm) with its $xy$ projection (gray). Thin black lines show the asymptotic net, green lines the two asymptotic lines through the tip, and red and black dotted lines the grip line and the section $y=0$ (radii of curvature $R_2$ and $R_1$).
    (c)--(e)~$R_1/R_2 = 0.5$, 1, and 4 ($R=R_2$). (c-i)--(e-i)~Side views at $\delta=5$\,mm with the simulated top surface (solid) and the undeformed shape (dotted); arrows mark the motion of the tips. (c-ii)--(e-ii)~Lit (yellow), penumbra (light gray), and shadow (gray) regions predicted by isometry (left) and simulated out-of-plane displacement $u_z/\delta$ on the undeformed planform (right). Dashed lines bound the lit region.}
    \label{fig:1}
\end{figure}

In this letter, we demonstrate that the asymptotic lines of shell structures, i.e., curves of zero normal curvature~\cite{doCarmo2016} on their mid-surface, dictate deformation and stiffness through pinching experiments on saddle-shaped shells with negative Gauss curvature, combining experiments, simulations, and theory (Fig.~\ref{fig:1}(a, b)). The isometry of their mid-surface yields the wave equation in space, from which the deformation of the saddle-shaped shells can be explained diagrammatically by identifying the lit, penumbra, and shadow regions.
Asymptotic lines characterize the rigidity landscape of the saddle-shell, which is nearly constant within the lit region and decays across the boundary into the shadow.
Our findings unify and complement the understanding of the localized deformation in pinching pipes and open cylindrical shells~\cite{Mahadevan2007, Matsumoto2018}, the stiffness of dome-like shells $(K_{\rm G}>0)$~\cite{Vella2012, SunPaulose2021} and of shells of mixed type~\cite{AudolyPomeau2002}, as well as provide the predictive foundation for the stiffness of the saddle-like geometry~\cite{Efrati2009} appearing in nature~\cite{LiangMahadevan2009, Dervaux2008, Nguyen2017, Zhang2025} or man-made structures~\cite{Klein2007, Kim2012, Gladman2016, Siefert2019}.

We start from pinching a saddle-shaped shell of thickness $h$, the shape of a \textit{potato chip}~\cite{Dervaux2008}, whose reference mid-surface, $\rf{z}=\rf{z}(x,y)$, is a hyperbolic paraboloid $\rf{z}(x,y)= x^2/2R_1 - y^2/2R_2$ with Gauss curvature $K_{\rm G}=-1/(R_1R_2)<0$, cut to a disk of radius $R$. The experimental saddle-shell sample is 3D-printed (Objet350 Connex3, Stratasys, USA) from resin (VeroClear, RGD810) with $(h, R, R_2)=(0.6, 20, 20)$~mm being fixed throughout. The radius of curvature along $x$, $R_1$, is chosen to range $R_1 = 5$--$80$~mm. The saddle-shell sits on a custom-made acrylic base that fits the curved geometry of the bottom tip; its compressive displacement (along the $y$-axis) is controlled by a micrometer (MT1/M, Thorlabs) so the shell is pinched along the grip-line $(x=0)$ by $\delta$ (Fig.~\ref{fig:1}(a)(b)). The deformed shape of the shell is recorded by the digital camera from $y<0$ as in Fig.~\ref{fig:1}(c) to measure the tip displacement $u_{\rm tip}$.
The tip falls when $R_1$ is below a critical radius $R_1^*$ and rises otherwise, and it barely moves at $R_1=R_2$, indicating $R_1^*\sim R_2$ (Fig.~\ref{fig:1}(c)).
We also simulate a discrete elastic shell~\cite{Grinspun2003, Grinspun2006}, where the edge points $(0,\pm R)$ are displaced toward each other by $\delta/2$ each~\cite{supp}.\nocite{BoydYu2011, Koiter1966, Underwood1983, Hinton1976} The simulated top-surface silhouettes agree well with the experimental photos (Fig.~\ref{fig:1}(c-i)--(e-i)).

\begin{figure}[!h]
    \centering
    \includegraphics[width=\linewidth]{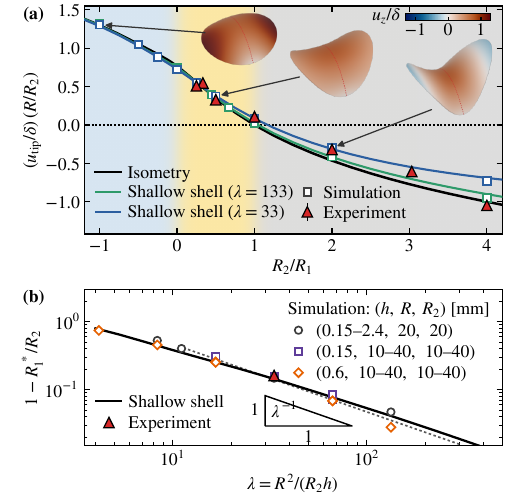}
    \caption{The relation between tip displacement and shell geometry. (a)~The empty squares and filled triangles represent simulation and experimental data. The black solid line represents the isometric prediction, whereas the colored curves show the numerical shallow shell theory solution. The colors represent the data and prediction for different thickness $\lambda$. The insets represent the deformed shape of the shells whose out-of-plane displacements are superimposed on the surface.
    (b)~The relation between the critical radius $R_1^*$ for the reversal of the tip displacement and shell thickness. The solid and dotted curves represent predictions from shallow shell theory and the scaling prediction. }
    \label{fig:2}
\end{figure}

The switching mechanism between rising and falling tips can be captured by the isometric bending limit, where the thickness is set to be exactly zero $h=0$.  Let $w=w(x,y)$ be the out-of-plane displacement of the mid-surface. The compatibility relation for the inextensible strain for shallow-shell theory becomes
\begin{equation}
    [\rf{z}, w] = \frac{1}{R_1}w_{,yy} - \frac{1}{R_2}w_{,xx} =0,
    \label{eq:MA_zw}
\end{equation}
where the bracket operator is the Monge--Amp\`ere bracket defined as $[f,g]\equiv f_{,xx}g_{,yy} + f_{,yy}g_{,xx} -2f_{,xy}g_{,xy}$ with the partial derivative denoted as $f_{,\mu} \equiv \partial_{\mu}f~(\mu=x,y)$. Equation~\eqref{eq:MA_zw} indicates that the out-of-plane displacement obeys the wave equation, with $x$ interpreted as time and wave speed $c = \sqrt{R_2/R_1}$. The general solution of Equation~\eqref{eq:MA_zw} is given by d'Alembert's formula, $w(x,y) = F(\xi) + F(\eta)$ with $\xi=y+cx$ and $\eta=y-cx$, where the even function $F=F(\xi)$ is to be derived to be consistent with the shape of the grip line, $w(0,y)=2F(y)$. In other words, the deformation along the grip line propagates along the asymptotic lines $y\pm cx=\mathrm{const}$. This isometric model divides the mid-surface into three regions according to how many of the two asymptotic lines through a point reach the grip line. Both reach it in the \textit{lit} region, only one in the \textit{penumbra}, and neither in the \textit{shadow}.
These regions (Fig.~\ref{fig:1}(c-ii)--(e-ii), left) classify the isometric response. The grip sets an upward deflection in the lit region, bending selects a downward one in the unconstrained shadow, and the penumbra superposes the two. The simulated displacement at the experimental thickness (right panels) shows that thickness blurs these boundaries.

The deformed configuration in the isometric limit is realized by minimizing the bending energy $U_{\rm b} = (D/2)\int\{(\Delta w)^2 - (1-\nu)[w,w]\} dA$ with the bending modulus $D\equiv Eh^3/\{12(1-\nu^2)\}$, Young's modulus $E$ and Poisson's ratio $\nu$, which is rewritten with respect to $g= F''(\xi)$ as
\begin{equation}
    U_{\rm b} = \frac{D}{2}\int\{(1+c^2)^2(g(\xi)+g(\eta))^2 - 8(1-\nu)c^2 g(\xi)g(\eta)\} dA,
\end{equation}
under the inextensible constraint of the grip line $\int_{-R} ^{R} w(0,y)dy = \delta R_2$ and the zero displacement condition at the pinch as $w(0,\pm R) = 0$. The isometric solution captures the trend of $u_{\rm tip}$ with $R_2/R_1$ and predicts the reversal exactly at $R_1=R_2$, but the simulations and experiments deviate slightly from it (Fig.~\ref{fig:2}(a)), indicating an effect of finite thickness.
The thickness allows the in-plane stretching of the mid-plane, whose governing equation is now given by the linearized shallow shell equation (the Marguerre-Donnell-Mushtari-Vlasov equation)
\begin{equation}
    D\nabla^4w - [\rf{z}, \Phi] =0, \quad \nabla^4\Phi + Eh[\rf{z}, w] = 0,
    \label{eq:MDMV}
\end{equation}
where the Airy stress function is introduced as $\Phi=\Phi(x,y)$, whose second derivatives yield the membrane stress as $N_{xx}=\Phi_{,yy}, N_{yy}=\Phi_{,xx}$, and $N_{xy}=-\Phi_{,xy}$~\cite{Marguerre1938, Donnell1933, Vlasov1964, supp}. Equation~\eqref{eq:MDMV} is summarized into a single PDE for $w$ as $D\nabla^8 w + Eh[\rf{z},[\rf{z}, w]] = 0$. By normalizing the lengths by $R$ and $w$ by $\delta R_2/R$, we find that the relevant dimensionless parameters are $R_1/R_2$,  $\lambda\equiv R^2/(R_2 h)$ and the Poisson's ratio $\nu$. Hence, the tip displacement, $u_{\rm tip}$, and the critical radius for tip rise or fall, $R_1^*$, are expected to satisfy the scaling function as
\begin{equation}
    u_{\rm tip} = \delta \frac{R_2}{R}\mathcal{G}\left(\frac{R_1}{R_2}; \lambda, \nu\right), \quad R_1^* = R_2\mathcal{F}(\lambda, \nu),
    \label{eq:scaling}
\end{equation}
with the dimensionless scaling functions, $\mathcal{G}$ and $\mathcal{F}$, which converge to the isometric solution as $\lambda\to\infty$. The shapes of the scaling functions, $\mathcal{G}$ and $\mathcal{F}$, are numerically obtained by minimizing the sum of the stretching and bending energy $U=U_{\rm b}+U_{\rm s}$, with the stretching energy defined as $U_{\rm s} = \int\{(\Delta \Phi)^2 - (1+\nu)[\Phi,\Phi]\} dxdy/(2Eh)$.
The minimization of the total elastic energy yields the scaling function for $u_{\rm tip}$, which is found to be in excellent agreement with the simulations in Fig.~\ref{fig:2}(a) ($\nu=0.4$ is fixed throughout). As the shell thickens, the $u_{\rm tip}$ curve shifts to the right, meaning the critical radius for the tip rise/fall, $R_1^*$, depends on $\lambda$. Indeed, the critical radius increases with $\lambda$ toward the isometric limit $R_1^* = R_2$ as shown in Fig.~\ref{fig:2}(b), whose scaling estimate is $1-R_1^*/R_2\sim \lambda^{-1}$.

\begin{figure}[!h]
    \centering
    \includegraphics[width=\linewidth]{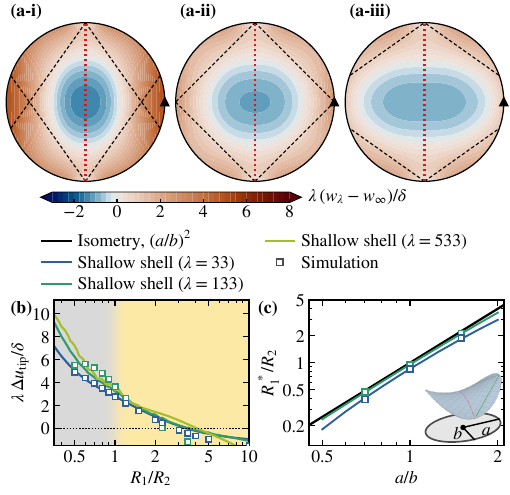}
    \caption{The effect of thickness on the out-of-plane displacement.
    (a)~The rescaled out-of-plane displacement fields relative to the prediction of the isometric deformation, $w_{\infty}$, are superposed onto the projected reference configurations obtained from the numerical solution of shallow shell theory with (a-i)~$R_1/R_2=0.5$, (a-ii)~$R_1/R_2=1.0$, and (a-iii)~$R_1/R_2=2.0$. The black dashed lines are the boundary between lit and shadow regions.
    (b) The scaling collapse of the relative tip displacement $\Delta u_{\rm tip}\equiv u_{\rm tip}- u_{\rm tip} ^{\rm iso}$ for several shell thicknesses.
    (c) The critical radius $R_1^*$ for the tip reversal in the case of an elliptical base of long and short axes $(a,b)$. (b)(c) The black and colored solid lines represent the isometric theory and shallow shell predictions, respectively.
    }
    \label{fig:3}
\end{figure}

\begin{figure}[!b]
    \centering
    \includegraphics[width=\linewidth]{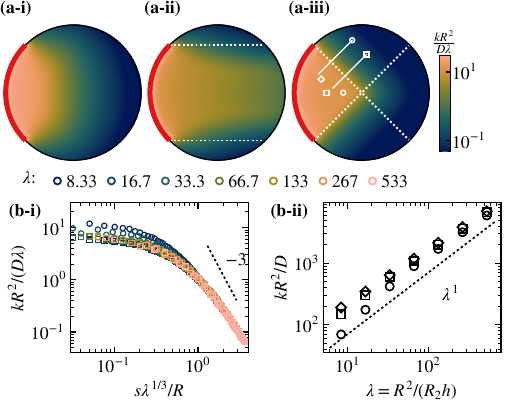}
    \caption{Rigidity landscape of the edge-clamped shell.
    (a) Heat map of the rescaled stiffness $kR^2/D\lambda$ for the (a-i) dome $R_2/R_1 = -1$, (a-ii) cylindrical $R_2/R_1 = 0$, and (a-iii) hyperbolic shells $R_2/R_1 = 1$ superposed onto the projected reference shape. The red arc ($3\pi/4$--$5\pi/4$) represents the clamped edge. The white dotted lines represent the asymptotic lines that cross two ends of the clamped edges.
    (b) Scaling collapse of the rigidity landscape. (b-i) Rescaled stiffness profiles normal to the lit--shadow boundary against $s\lambda^{1/3}/R$, with $s$ the distance into the shadow, along the lines in (a-iii) (empty circles and squares) for several $\lambda$. The dashed line shows $k\sim DR/s^3$.
    (b-ii) The rescaled stiffness on the lit region measured at the location denoted by empty diamonds, squares, and circles in (a-iii). The dashed line represents the scaling prediction as $k\sim D\lambda/R^2$. }
    \label{fig:4}
\end{figure}

The rise and fall of the tip is characteristic of the saddle-shaped shell. The numerical minimization techniques can be employed to the dome-like and cylindrical shell $(R_2/R_1\leq0)$, whose results are smoothly connected to those of the saddle-shaped shell as shown in Fig.~\ref{fig:2}(a). We find that the tip always rises, $u_{\rm tip} > 0$, for the $K_{\rm G}>0$ shell, emphasizing the role of asymptotic lines. We further clarify this role by utilizing the numerically obtained solutions of shallow shell theory. We plot the heat map of the out-of-plane displacement, $w_{\lambda}$, measured from the isometric prediction, $w_{\infty}$, rescaled as $\lambda(w_{\lambda}-w_{\infty})/\delta$ in Fig.~\ref{fig:3}(a), where the asymptotic lines passing through the pinching points are plotted as black dashed lines. We note that the thickness correction to the isometric displacement, $w_{\lambda} = w_{\infty} + \lambda^{-1}w_1 + O(\lambda^{-2})$, in the lit region (i.e., the diamond regions enclosed by the asymptotic lines) is small, whereas it is large outside the lit region (Fig.~\ref{fig:3}(a-i), near the tip). This difference is illustrated in the rescaled tip displacement, $\lambda \Delta u_{\rm tip}/\delta$, against $R_1/R_2$ collapsed onto the single master curve in Fig.~\ref{fig:3}(b). Changing the planform to an ellipse with long and short axes $(a,b)$ confirms that the geometry of asymptotic lines matters. The asymptotic lines through the pinching points predict $R_1^*/R_2=(a/b)^2$ in the isometric limit, and $R_1^*$ decreases with thickness (Fig.~\ref{fig:3}(c)).

We now illustrate that the asymptotic line also contributes to the rigidity of the shell. We clamp a quarter of the planform edge (shown as a red arc in Fig.~\ref{fig:4}(a)) and indent a shell to measure the local rigidity of the shell, $k$. To compute the rigidity landscape efficiently, we extract the rigidity matrix of the discrete elastic shell under clamped boundary conditions~\cite{supp}. We plot the rigidity heatmaps normalized as $kR^2/D\lambda$ for dome-like, cylindrical, and saddle-shaped shells in Fig.~\ref{fig:4}(a-i), (a-ii), and (a-iii), respectively, finding the unique feature of the saddle-shaped shell.
The saddle-shaped shell is stiff within a diamond bounded by the two asymptotic lines through the ends of the clamped arc. This diamond is the region lit by the clamp, and the local stretch (change in Gauss curvature) induced by indentation spreads only within it, as \textit{illuminated} in Fig.~\ref{fig:4}(a-iii). The stiffness is nearly constant there and decays across the lit--shadow boundary (Fig.~\ref{fig:4}(b-i)).
Let $\ell$ be the characteristic deformation size, with a bending penalty of $\sim D(w/\ell^2)^2\ell^2$ for the out-of-plane displacement $w$. The in-plane stretch accounted for by the indentation is $\sim w/R_2$, from which the stretching energy is scaled as $Eh(w/R_2)^2\ell^2$. Balancing stretching and bending gives the characteristic length scale $\ell\sim\sqrt{hR_2}$, so the stiffness scales as $k\sim Eh^2/R_2\sim D\lambda/R^2$, consistent with the simulation as shown in Fig.~\ref{fig:4}(b-ii).
Outside the lit region, isometric bending is allowed; i.e., the rigidity decays irrespective of $\lambda$ as $k\sim DR/s^3$, with $s$ the distance from the lit--shadow boundary, consistent with the simulation results in Fig.~\ref{fig:4}(b-i). The two regimes are joined by a boundary layer along the asymptotic lines. Across an asymptotic line, the operator $[\rf{z},\,\cdot\,]$ involves only one derivative normal to the line, so that for a layer of width $s$ and length $\sim R$, the single PDE $D\nabla^8 w + Eh[\rf{z},[\rf{z},w]]=0$ balances as $D/s^8\sim Eh/(R_2 sR)^2$, giving $s\sim R\lambda^{-1/3}\propto h^{1/3}$. Accordingly, the profiles in Fig.~\ref{fig:4}(b-i) collapse onto $kR^2/(D\lambda)=f(s\lambda^{1/3}/R)$, whose tail $f(u)\sim u^{-3}$ recovers $k\sim DR/s^3$.
Such confined stiffness is absent for the dome ($K_{\rm G}>0$, Fig.~\ref{fig:4}(a-i)), whereas the cylinder ($K_{\rm G}=0$, Fig.~\ref{fig:4}(a-ii)) is the degenerate case, with a stiff band bounded by its generators. The exponents agree with the asymptotic theory of shells~\cite{Pitkaranta2001, Karamian2002, AudolyPomeau2002}. The light-and-shadow rule adds that these layers sit on the asymptotic lines through the ends of the support.

We have shown that the asymptotic lines of a saddle-shaped shell act as rays that carry boundary constraints into its surface. Isometry turns the mid-surface into a medium obeying a wave equation in space, and the grip line divides it into lit, penumbra, and shadow regions. The lit region rises with the pinch, whereas the shadow, unconstrained by the grip, sinks.
The tip therefore reverses at $R_1^*=R_2$ for a disk and at $(a/b)^2R_2$ for an ellipse, a criterion set by where the asymptotic lines reach rather than by the local curvature; finite thickness blurs the lit--shadow boundary at order $\lambda^{-1}$. In a clamped shell, the same rule separates a lit region, where indentation requires stretching, from a shadow, where bending alone sets the stiffness. Dome-like shells, whose isometry obeys an elliptic equation, cast no shadows, and cylinders recover the persistent pinch of pipes and open cylindrical shells~\cite{Mahadevan2007, Matsumoto2018}.
Deformation of a thin shell propagates along the characteristics of a hyperbolic equation, as do stresses in granular media~\cite{Cates1998, Geng2001} and mechanical metamaterials~\cite{Zheng2022}; here, however, the medium is the surface geometry itself.
Because saddle-like curvature is ubiquitous in nature and can now be programmed~\cite{LiangMahadevan2009, Kim2012, Siefert2019}, light and shadow offer a simple way to anticipate where a curved sheet will move and where it will hold, from stretchable electronics to deployable structures~\cite{Rogers2010, Melancon2021}. In a shell, curvature not only stores shape; it also determines where mechanical information can travel.

The same geometry may also guide failure. The stress of a pinched shell concentrates in the lit region while the shadow remains nearly unloaded~\cite{supp}, and flexural waves on saddles travel close to the asymptotic directions~\cite{EvansLevine2013}, suggesting that, as in the cascading fracture of spaghetti~\cite{AudolyNeukirch2005}, cracks may be channeled along the lit--shadow boundary.

\begin{acknowledgments}
This work was supported by MEXT KAKENHI 24H00299 and JST FOREST Program, Grant Number JPMJFR212W. 
\end{acknowledgments}

\bibliography{apssamp}

\clearpage
\onecolumngrid
\setcounter{secnumdepth}{3}
\setcounter{section}{0}
\setcounter{equation}{0}
\setcounter{figure}{0}
\renewcommand{\thefigure}{S\arabic{figure}}
\renewcommand{\theequation}{S\arabic{equation}}

\begin{center}
{\large\bfseries Supplemental Material for ``Pinching an elastic potato chip''}\\[8pt]
Kosei Kataoka and Tomohiko G. Sano\\[2pt]
{\small\itshape Department of Mechanical Engineering, Keio University, Yokohama, Kanagawa 223-8522, Japan}
\end{center}
\vspace{4pt}
\noindent{\small Equation and figure numbers without the prefix S, and all reference numbers, refer to the main text above.}
\vspace{8pt}

\section{Marguerre-Donnell-Mushtari-Vlasov equation}\label{sec:MDMVeq}

We review the theoretical basis of our manuscript, formulated as the Marguerre--Donnell--Mushtari--Vlasov equation~\cite{Marguerre1938, Donnell1933, Vlasov1964}. We start describing elastic stretching and bending energies in Sec.~\ref{sec:en}. The compatibility relation follows in Sec.~\ref{sec:comp}, whose difference from the Donnell--Mushtari--Vlasov equation, formulation for constant curvature, is detailed. Using the principle of virtual work, we derive the Marguerre--Donnell--Mushtari--Vlasov equation used in the main text in Sec.~\ref{sec:pvw}. The linearized version follows in Sec.~\ref{sec:lin} and numerical procedures to obtain solutions are detailed for isometric and finite thickness cases in Secs.~\ref{sec:iso} and \ref{sec:finite}, respectively.

\subsection{Elastic energy: Stretching and Bending}\label{sec:en}

We describe the reference mid-surface of a shallow shell as the surface $z=\rf{z}(x,y)$ over the $xy$ plane, assuming the reference shape is shallow, i.e., $|\nabla\rf{z}|\ll1$. The mid-surface is displaced by the displacement vector $(u_x,u_y,w)=(u,v,w)$. Greek indices are used for $x$ and $y$, and repeated indices are summed $(\alpha,\beta=x,y)$. Retaining the leading terms in the slopes, the change in length of a line element of the mid-surface gives the membrane strain~\cite{Marguerre1938}, and the change in curvature is given by the second derivatives of $w$:
\begin{equation}
    \varepsilon_{\alpha\beta}=\frac{1}{2}\left(u_{\alpha,\beta}+u_{\beta,\alpha}\right)+\frac{1}{2}\left(\rf{z}_{,\alpha}w_{,\beta}+\rf{z}_{,\beta}w_{,\alpha}\right)+\frac{1}{2}w_{,\alpha}w_{,\beta},
    \qquad \kappa_{\alpha\beta}=w_{,\alpha\beta}.
    \label{eq:m_strain}
\end{equation}
As in the main text, the partial derivative is denoted as $f_{,\mu} = \partial_{\mu}f$.
Compared with a flat plate, the only new ingredient is the second term of $\varepsilon_{\alpha\beta}$: on a curved surface, an out-of-plane displacement stretches the mid-surface already at linear order. The stretching and bending energies of an isotropic elastic shell are
\begin{align}
    U_{\rm s}&=\frac{Eh}{2(1-\nu^2)}\int\left[(1-\nu)\varepsilon_{\alpha\beta}\varepsilon_{\alpha\beta}+\nu\,\varepsilon_{\alpha\alpha}\varepsilon_{\beta\beta}\right]dA, \nonumber\\
    U_{\rm b}&=\frac{D}{2}\int\left[(1-\nu)\kappa_{\alpha\beta}\kappa_{\alpha\beta}+\nu\,\kappa_{\alpha\alpha}\kappa_{\beta\beta}\right]dA=\frac{D}{2}\int\left\{(\nabla^2w)^2-(1-\nu)[w,w]\right\}dA,
    \label{eq:m_energy}
\end{align}
with $D=Eh^3/\{12(1-\nu^2)\}$. In components, $U_{\rm s}$ is the expression given in the main text, with $\gamma_{xy}=2\varepsilon_{xy}$.

\subsection{Compatibility relation}\label{sec:comp}

The three strain components derive from the two in-plane displacements and are therefore not independent. Eliminating $u_\alpha$ from the expression of strain tensors, Eq.~\eqref{eq:m_strain} yields the compatibility relation
\begin{equation}
    \varepsilon_{xx,yy}+\varepsilon_{yy,xx}-\gamma_{xy,xy}=-[\rf{z},w]-\frac{1}{2}[w,w],
    \label{eq:m_comp}
\end{equation}
indicating that the in-plane strain and the out-of-plane displacement are coupled.
Given that the Gauss curvature of a shallow surface $z$ is denoted with the Monge--Amp\`ere bracket as $[z,z]/2$, the right-hand side is minus the change in the Gauss curvature, $[\rf{z}+w,\rf{z}+w]/2-[\rf{z},\rf{z}]/2=[\rf{z},w]+[w,w]/2$.
In particular, an isometric deformation, $\varepsilon_{\alpha\beta}=0$, satisfies $[\rf{z},w]=0$ at linear order, which is Eq.~(1) of the main text.

The bracket $[\rf{z},\,\cdot\,]$ is where Marguerre's formulation~\cite{Marguerre1938} differs from the Donnell--Mushtari--Vlasov (DMV) equations familiar for spherical and cylindrical shells~\cite{Donnell1933, Vlasov1964}. The DMV theory is formulated for a shell with constant principal curvatures $1/R_x$ and $1/R_y$ along the coordinate lines, i.e., $\rf{z}=x^2/(2R_x)+y^2/(2R_y)$, for which the bracket reduces to the Vlasov operator,
\begin{equation}
    [\rf{z},f]=\frac{1}{R_x}f_{,yy}+\frac{1}{R_y}f_{,xx},
    \label{eq:m_vlasov}
\end{equation}
and, for a spherical shell of radius $R_x=R_y=R_{\rm s}$, to the Laplacian, $\nabla^2f/R_{\rm s}$. Marguerre's form holds for an arbitrary shallow reference shape, including varying curvature and twist ($\rf{z}_{,xy}\neq0$), and makes the type of the geometric coupling explicit: the principal part of $[\rf{z},\,\cdot\,]$ has the determinant $\rf{z}_{,xx}\rf{z}_{,yy}-\rf{z}_{,xy}^2\simeq K_{\rm G}$, so that the operator is elliptic for a dome, parabolic for a cylinder, and hyperbolic for a saddle. For a spherical shell, isometric deformations are harmonic, $\nabla^2w=0$, and have no asymptotic lines. For the hyperbolic paraboloid of the main text, $R_x=R_1$ and $R_y=-R_2$, and $[\rf{z},\,\cdot\,]$ becomes the wave operator of Eq.~(1) of the main text, whose characteristics are the asymptotic lines.
Given that the second derivatives of this $\rf{z}$ are constant, our equations are equivalent to the DMV equations with principal radii $R_1$ and $-R_2$; we use the Marguerre form because it treats domes, cylinders, and saddles in a single expression.

\subsection{Principle of virtual work}\label{sec:pvw}

The equilibrium follows from the principle of virtual work. In the absence of body forces, the first variation of $U=U_{\rm s}+U_{\rm b}$ vanishes for any variation $(\delta u_\alpha,\delta w)$ compatible with the constraints at the grip points. The variations of Eqs.~\eqref{eq:m_strain} and \eqref{eq:m_energy} yield
\begin{equation}
    \delta U=\int\left[N_{\alpha\beta}\,\delta u_{\alpha,\beta}+N_{\alpha\beta}(\rf{z}+w)_{,\alpha}\,\delta w_{,\beta}+M_{\alpha\beta}\,\delta w_{,\alpha\beta}\right]dA,
    \label{eq:m_pvw}
\end{equation}
where $N_{\alpha\beta}=\{Eh/(1-\nu^2)\}[(1-\nu)\varepsilon_{\alpha\beta}+\nu\varepsilon_{\gamma\gamma}\delta_{\alpha\beta}]$ and $M_{\alpha\beta}=D[(1-\nu)\kappa_{\alpha\beta}+\nu\kappa_{\gamma\gamma}\delta_{\alpha\beta}]$ are the membrane force and the bending moment. After integration by parts, the bulk terms give the in-plane equilibrium, $N_{\alpha\beta,\beta}=0$, and the out-of-plane equilibrium, $D\nabla^4w-N_{\alpha\beta}(\rf{z}+w)_{,\alpha\beta}=0$, whereas the boundary terms give the vanishing edge forces and moment conditions. The in-plane equilibrium is satisfied automatically by the Airy stress function $\Phi$, $N_{xx}=\Phi_{,yy}$, $N_{yy}=\Phi_{,xx}$, and $N_{xy}=-\Phi_{,xy}$. With this Airy stress representation, the normal component of the membrane force acting on a surface of height $f = \rf{z} + w$ is written as $N_{\alpha\beta}f_{,\alpha\beta}=[\Phi,f]$.
Expressing the strains in Eq.~\eqref{eq:m_comp} through $\Phi$, we obtain the Marguerre--DMV equations,
\begin{equation}
    D\nabla^4w-[\rf{z}+w,\Phi]=0,\qquad \nabla^4\Phi+Eh\left([\rf{z},w]+\frac{1}{2}[w,w]\right)=0.
    \label{eq:m_MDMV}
\end{equation}
For a flat plate, $\rf{z}=0$, Eq.~\eqref{eq:m_MDMV} reduces to the F\"oppl--von K\'arm\'an equations.

\subsection{Linearized equation}\label{sec:lin}

When the curvature change is much smaller than the reference curvature, $|w_{,\alpha\beta}|\ll1/R_{1},1/R_{2}$, the nonlinear terms $[w,\Phi]$ and $[w,w]$ in Eq.~\eqref{eq:m_MDMV} are negligible compared with $[\rf{z},\Phi]$ and $[\rf{z},w]$, and Eq.~\eqref{eq:m_MDMV} reduces to
\begin{equation}
    D\nabla^4w-[\rf{z},\Phi]=0,\qquad \nabla^4\Phi+Eh[\rf{z},w]=0,
    \label{eq:m_lin}
\end{equation}
which is Eq.~(3) of the main text. When the second derivatives of $\rf{z}$ are constant, as for the quadric surfaces considered here, $[\rf{z},\,\cdot\,]$ commutes with $\nabla^4$, and eliminating $\Phi$ gives the single equation $D\nabla^8w+Eh[\rf{z},[\rf{z},w]]=0$. Equation~\eqref{eq:m_lin} can also be derived directly from the principle of virtual work by omitting the nonlinear term $w_{,\alpha}w_{,\beta}/2$ from Eq.~\eqref{eq:m_strain}. The energy $U$ is quadratic in $(u,v,w)$, and Eq.~\eqref{eq:m_lin} is its stationary condition. In Secs.~\ref{sec:iso} and \ref{sec:finite}, we obtain the solutions by minimizing this quadratic energy instead of solving Eq.~\eqref{eq:m_lin} directly.
Measuring lengths in units of $R$, $w$ in units of $\delta R_2/R$, and $(u,v)$ in units of $\delta$, the energy in units of $D\delta^2R_2^2/R^4$ reads
\begin{equation}
    \tilde{U}=\frac{1}{2}\int\left\{(\nabla^2w)^2-(1-\nu)[w,w]\right\}dA+6\lambda^2\int\left[\varepsilon_{xx}^2+\varepsilon_{yy}^2+2\nu\varepsilon_{xx}\varepsilon_{yy}+\frac{1-\nu}{2}\gamma_{xy}^2\right]dA,
    \label{eq:m_nondim}
\end{equation}
where the strains contain the rescaled reference surface $\rf{z}R_2/R^2=x^2/(2R_1/R_2)-y^2/2$ and $\lambda=R^2/(R_2h)$. The solution thus depends only on $R_1/R_2$, $\lambda$, and $\nu$, as in Eq.~(4) of the main text.

\subsection{Numerical procedure to obtain isometric solutions}\label{sec:iso}

In the isometric limit, $\lambda\to\infty$, the stretching term in Eq.~\eqref{eq:m_nondim} enforces $\varepsilon_{\alpha\beta}=0$: $w$ satisfies $[\rf{z},w]=0$ and minimizes $U_{\rm b}$. For the saddle, $w=F(\xi)+F(\eta)$ with $\xi,\eta=y\pm cx$ and $c=\sqrt{R_2/R_1}$, and $F$ is even by symmetry. For the dome ($R_1<0$), $[\rf{z},w]=0$ is elliptic, and $w=F(y+icx)+F(y-icx)=2\,\mathrm{Re}\,F(y+icx)$ with $c=\sqrt{R_2/|R_1|}$; the cylinder corresponds to $c=0$. In all cases, $w(0,y)=2F(y)$. The pinch enters through the grip line: integrating the inextensibility along $x=0$, $\varepsilon_{yy}=v_{,y}-(y/R_2)w_{,y}=0$, over $-R\leq y\leq R$ with $v(0,\pm R)=\mp\delta/2$ and $w(0,\pm R)=0$ yields the constraints of the main text,
\begin{equation}
    \int_{-R}^{R}w(0,y)\,dy=\delta R_2,\qquad F(\pm R)=0.
    \label{eq:m_grip}
\end{equation}

The numerical isometric solutions are obtained as follows. We expand $g=F''$ in piecewise-linear functions on $0\leq\xi\leq R\sqrt{1+c^2}$ (400 elements). As the bending energy depends solely on $g$, it is the unknown of the minimization, and $F$ is recovered by integrating it twice as $F(\xi)=\int_0^{\xi}(\xi-s)g(s)\,ds-\int_0^{R}(R-s)g(s)\,ds$, which satisfies $F'(0)=0$ (symmetry) and $F(R)=0$. The bending energy, Eq.~(2) of the main text, becomes a quadratic form $\bm{g}^{\rm T}\bm{H}\bm{g}/2$ in the nodal values $\bm{g}$, where $\bm{H}$ is the Hessian matrix of $U_{\rm b}$ with respect to $\bm{g}$, evaluated by quadrature on a uniform grid over the disk, and substituting $F$ into Eq.~\eqref{eq:m_grip}, or equivalently integrating $\int F\,dy$ twice by parts with $F(\pm R)=0$, turns the grip constraint into a linear one in $g$,
\begin{equation}
    \int_{-R}^{R}(R^2-s^2)\,g(s)\,ds=-\delta R_2.
    \label{eq:m_gripg}
\end{equation}
Minimizing the quadratic form under this constraint with a Lagrange multiplier $\mu$ yields the linear equation $\bm{H}\bm{g}=\mu\bm{a}$, where $\bm{a}^{\rm T}\bm{g}$ is the left-hand side of Eq.~\eqref{eq:m_gripg}, and hence $\bm{g}=\mu\bm{H}^{-1}\bm{a}$ with $\mu$ fixed by the constraint.
The numerical isometric solution for the dome is computed as follows. We expand $F$ in even polynomials, $F(\zeta)=\sum_{k=0}^{14}a_k\zeta^{2k}$, so that $w$ is a sum of harmonic polynomials, and minimize $U_{\rm b}$ under Eq.~\eqref{eq:m_grip} in the same way.

These isometric solutions for domes, cylinder, and saddles agree with the finite-thickness solutions of Sec.~\ref{sec:finite} at $\lambda=10^4$ to within $5\times10^{-3}$ in $u_{\rm tip}R/(\delta R_2)$ for $-2\leq R_2/R_1\leq4$; in Figs.~2(a) and 3(a, b) of the main text, the isometric predictions are evaluated with the latter.

\subsection{Numerical procedure to obtain finite thickness cases}\label{sec:finite}

For finite $\lambda$, we minimize the energy of Eq.~\eqref{eq:m_nondim} with a spectral Galerkin method~\cite{BoydYu2011}. The displacements are expanded in Zernike polynomials on the unit disk, in the polar coordinates $(r,\theta)$ with $x=Rr\cos\theta$ and $y=Rr\sin\theta$, up to the total degree $N=56$, with the symmetry of the problem ($w$ even in $x$ and $y$, $u$ odd in $x$, and $v$ odd in $y$),
\begin{equation}
    w=\sum_{n,\,m\,{\rm even}}a_{nm}R_n^m(r)\cos m\theta,\quad u=\sum_{n,\,m\,{\rm odd}}b_{nm}R_n^m(r)\cos m\theta,\quad v=\sum_{n,\,m\,{\rm odd}}c_{nm}R_n^m(r)\sin m\theta,
    \label{eq:m_zernike}
\end{equation}
where $R_n^m(r)=r^mP^{(0,m)}_{(n-m)/2}(2r^2-1)$ ($n\geq m$, $n-m$ even) are the radial Zernike polynomials, with $P^{(\alpha,\beta)}_k$ the Jacobi polynomials. The energy becomes a quadratic form, $\tilde{U}=\frac{1}{2}\bm{q}^{\rm T}\bm{K}\bm{q}$, in the coefficients $\bm{q}=(a_{nm},b_{nm},c_{nm})$, with $\bm{K}$ evaluated by Gauss--Legendre quadrature in $r$ and the trapezoidal rule in $\theta$, exact for the polynomials used. The pinch is imposed at the grip points by $w=0$ and $v=\mp\delta/2$ at $(0,\pm R)$.
By symmetry (Eq.~\eqref{eq:m_zernike}), one of these gripping-point conditions is sufficient.
Evaluating the expansions at $(0, R)$ gives a linear constraint $\bm{C}\bm{q}=\bm{d}$ with $\bm{d}=(0,-\delta/2)^{\rm T}$, so that
\begin{equation}
    \begin{pmatrix}\bm{K}&\bm{C}^{\rm T}\\ \bm{C}&\bm{0}\end{pmatrix}\begin{pmatrix}\bm{q}\\ \bm{\mu}\end{pmatrix}=\begin{pmatrix}\bm{0}\\ \bm{d}\end{pmatrix},
    \label{eq:m_galerkin}
\end{equation}
with the Lagrange multipliers $\bm{\mu}$. We impose no condition on the expansion at the edge $r=1$. The vanishing edge forces and moments conditions follow from the minimization as natural boundary conditions. The elliptic planform in Fig.~3(c) of the main text is mapped onto the unit disk by rescaling $x$ by $a/b$, with lengths in units of $b$. Changing $N$ between 48 and 64 changes $u_{\rm tip}$ only in the third decimal place of $u_{\rm tip}R/(\delta R_2)$. As $\lambda\to\infty$, the stretching term enforces $\varepsilon_{\alpha\beta}=0$, the point constraints reduce to Eq.~\eqref{eq:m_grip}, and the solution approaches the isometric one of Sec.~\ref{sec:iso}.

\section{Discrete Elastic Shells}\label{sec:DES}

All simulations in the main text are performed with a discrete elastic shell formalism, where the mid-surface of the shell is discretized into a set of triangles, whose stretching and bending energies are calculated via the strain of triangles and the relative angles between normal vectors of adjacent triangles~\cite{Grinspun2003, Grinspun2006}. This method is known to reproduce the mechanical performance of plate or shell structures, such as the Reissner stiffness. Although the protocol has been proposed and validated in the previous work~\cite{Grinspun2003, Grinspun2006}, we summarize the simulation method employed in the main text for completeness. In Sec.~\ref{sec:overview}, we present the simulation formalism, including elastic stretching and bending energies. Next, in Sec.~\ref{sec:detail}, we derive the equations of motion for nodes of triangles based on the energy function in Sec.~\ref{sec:overview}. The simulation protocol and specific information in our setup are summarized in Sec.~\ref{sec:specific}. In Sec.~\ref{sec:validation}, we compare our discrete simulation algorithms against the conventional finite element method (FEM) to validate the formulations. In Sec.~\ref{sec:rigidity}, we describe the algorithms to compute the rigidity landscape.

\subsection{Formulation}\label{sec:overview}

\begin{figure}[!h]
    \centering
    \includegraphics[width=3.4in]{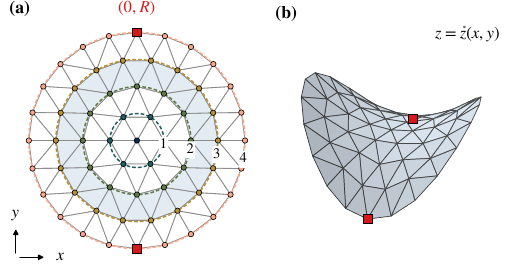}
    \caption{Reference mesh of the discrete elastic shell, illustrated with $N_{\rm r}=4$. (a)~Planform seen from above. A center node and $6k$ equally spaced nodes on the circle of radius $kR/N_{\rm r}$ (dashed; numbers: $k$) are connected by nearly equilateral triangles; the shaded band between the circles $k=2$ and 3 contains $6\times(2+3)$ triangles. Red squares: grip nodes $(0,\pm R)$. (b)~The nodes placed on the reference surface $z=\rf{z}(x,y)$ ($R_1=R_2$). The simulations use $N_{\rm r}=20$--$114$.}
    \label{fig:mesh}
\end{figure}

The reference mid-surface, $z=\rf{z}(x,y)$, is discretized into a set of triangles, whose vertices correspond to the nodes in our simulations. The nodes consist of a node at the center and $6k$ equally spaced nodes on each of the concentric circles of radius $kR/N_{\rm r}$ ($k=1,2,\cdots, N_{\rm r}$), and adjacent circles are connected by nearly equilateral triangles [Fig.~\ref{fig:mesh}(a)] (the $x$ coordinates are multiplied by $a/b$ for the elliptic planform in Fig.~3(c) of the main text). Each node is then placed on the reference surface, $\bm{X}_i=(X_i,Y_i,\rf{z}(X_i,Y_i))$ [Fig.~\ref{fig:mesh}(b)], and its current position $\bm{x}_i(t)$ evolves according to the equation of motion in Sec.~\ref{sec:detail}. The elastic energy is the sum of the stretching and bending energies, $U=U_{\rm s}+U_{\rm b}$.

The elastic energy in our simulation is the discretized version of the stretching and bending energies used in the continuum model (Eqs.~\eqref{eq:m_energy}).
The stretching energy is evaluated with constant-strain triangles of a plane-stress St.~Venant--Kirchhoff material,
\begin{equation}
    U_{\rm s}=\sum_{T}\frac{Eh\,A_T}{2(1-\nu^2)}\left[(\mathrm{tr}\,\bm{\varepsilon}_T)^2-2(1-\nu)\det\bm{\varepsilon}_T\right],
    \label{eq:Us}
\end{equation}
where $A_T$ is the reference area of triangle $T$. For a triangle with nodes $(i_0,i_1,i_2)$, we introduce an orthonormal frame on the reference triangle and express its reference edge vectors in this frame as the columns of a $2\times2$ matrix, $\bm{D}_{\rm m}=\left(\bm{X}_{i_1}-\bm{X}_{i_0}\,\middle|\,\bm{X}_{i_2}-\bm{X}_{i_0}\right)$. With the current edge vectors as columns, $\bm{D}_{\rm s}=\left(\bm{x}_{i_1}-\bm{x}_{i_0}\,\middle|\,\bm{x}_{i_2}-\bm{x}_{i_0}\right)$ ($3\times2$), the deformation gradient and the Green--Lagrange strain of the triangle are
\begin{equation}
    \bm{F}_T=\bm{D}_{\rm s}\bm{D}_{\rm m}^{-1},\qquad \bm{\varepsilon}_T=\frac{1}{2}\left(\bm{F}_T^{\rm T}\bm{F}_T-\bm{I}\right).
    \label{eq:strain}
\end{equation}
As $\bm{F}_T^{\rm T}\bm{F}_T$ depends only on the inner products of the edge vectors, the Green--Lagrange strain tensor $\bm{\varepsilon}_T$ is determined by the changes in the three edge lengths of the triangle and vanishes for rigid-body motions.

The bending energy takes the Koiter form~\cite{Koiter1966},
\begin{equation}
    U_{\rm b}=\sum_{T}\frac{D\,A_T}{2}\left[(\mathrm{tr}\,\Delta\bm{b}_T)^2-2(1-\nu)\det\Delta\bm{b}_T\right],
    \label{eq:Ub}
\end{equation}
where $\Delta\bm{b}_T$ is the change in the discrete shape operator of triangle $T$, representing the change of local curvature. Following the idea of Ref.~\cite{Grinspun2006}, the shape operator $\Delta\bm{b}_T$ is reconstructed from changes in the dihedral angles at the three edges of triangle $T$. The dihedral angle at an interior edge $e$ shared by triangles $A$ and $B$ is the signed angle between their unit normals $\bm{n}_A$ and $\bm{n}_B$ as $\theta_e\equiv\arctan[(\bm{n}_A\times\bm{n}_B)\cdot\hat{\bm{e}}/(\bm{n}_A\cdot\bm{n}_B)]$, with $\hat{\bm{e}}$ the unit vector along the edge. We write
\begin{equation}
    \Delta\bm{b}_T=\sum_{e\in T}\bm{C}_{T,e}\left(\theta_e-\bar{\theta}_e\right),
    \label{eq:Db}
\end{equation}
where $\bar{\theta}_e$ is the dihedral angle of the reference mesh, so that the undeformed mesh is stress-free and its natural curvature is that of the reference surface.
The constant tensors $\bm{C}_{T,e}$, that connect dihedral angle and curvature, are determined once from the reference mesh. Consider a quadratic surface over the plane of a triangle $T$, $z=b_{\alpha\beta}\xi_\alpha\xi_\beta/2$ with in-plane coordinates $(\xi_1,\xi_2)$, and place the vertices of $T$ and of its three neighbors on it. The vertex heights are linear in the three components $(b_{11},b_{22},b_{12})$, and so are the slopes $\nabla z_T$ and $\nabla z_{T'}$ of the two flat triangles $T$ and $T'$ sharing an edge $e$. The dihedral angle at $e$ is, to leading order in the small dihedral angles, the difference in these slopes normal to the edge, $\theta_e\simeq(\nabla z_T-\nabla z_{T'})\cdot\hat{\bm{e}}_\perp$, with $\hat{\bm{e}}_\perp$ the unit vector in the plane of $T$ perpendicular to the edge, and is therefore a linear combination of $b_{11}$, $b_{22}$, and $b_{12}$. Writing this for the three edges of the triangle $T$ gives
\begin{equation}
    \begin{pmatrix}\theta_1\\ \theta_2\\ \theta_3\end{pmatrix}=\bm{Q}_T\begin{pmatrix}b_{11}\\ b_{22}\\ b_{12}\end{pmatrix},
    \qquad
    \begin{pmatrix}b_{11}\\ b_{22}\\ b_{12}\end{pmatrix}=\bm{Q}_T^{-1}\begin{pmatrix}\theta_1\\ \theta_2\\ \theta_3\end{pmatrix},
    \label{eq:Mtheta}
\end{equation}
where the $3\times3$ matrix $\bm{Q}_T$ depends only on the vertex positions of the reference mesh. The $e$-th column of $\bm{Q}_T^{-1}$, arranged as a symmetric tensor, is $\bm{C}_{T,e}$. Equation~\eqref{eq:Db} thus reproduces the curvature change of any quadratic surface, up to corrections of relative order $\theta_e^2$ with $\theta_e\sim\ell_e/R_{1,2}$, which vanish as the mesh is refined. For an edge of $T$ lying on the boundary of the shell, where no neighboring triangle exists, the corresponding row of Eq.~\eqref{eq:Mtheta} is replaced by the condition of vanishing normal bending moment at the free edge, $\hat{\bm{e}}_\perp\cdot\bm{b}\cdot\hat{\bm{e}}_\perp+\{\nu/(1-\nu)\}\,\mathrm{tr}\,\bm{b}=0$.
Both energies retain the Poisson-ratio dependence of their continuum counterparts in the main text. Lengths are measured in mm and energies in units of $D$. As the shell is loaded by prescribed displacements, the equilibrium shape does not depend on the value of the Young's modulus $E$. We fix Poisson's ratio $\nu=0.4$ throughout.

\subsection{Equation of motion}\label{sec:detail}

As we are interested only in equilibrium shapes, we obtain them by dynamic relaxation~\cite{Underwood1983}: the position of each node evolves according to the damped equations of motion,
\begin{equation}
    m\frac{d^2\bm{x}_i}{dt^2}=\bm{F}_i^{\rm s}+\bm{F}_i^{\rm b}-m\gamma\frac{d\bm{x}_i}{dt},
    \qquad \bm{F}_i^{\rm s}=-\frac{\partial U_{\rm s}}{\partial\bm{x}_i},\quad \bm{F}_i^{\rm b}=-\frac{\partial U_{\rm b}}{\partial\bm{x}_i},
    \label{eq:eom}
\end{equation}
with a numerical nodal mass $m$ and a damping rate $\gamma$; neither affects the equilibria.

The stretching force on node $i$, $\bm{F}_i^{\rm s}=-\partial U_{\rm s}/\partial\bm{x}_i$, is the sum of the contributions of all triangles sharing node $i$. Writing Eq.~\eqref{eq:Us} as $U_{\rm s}=\sum_TA_TW_T$, with $W_T$ the energy per unit reference area of triangle $T$, the energy of $T$ depends on its nodes only through the deformation gradient $\bm{F}_T$ in Eq.~\eqref{eq:strain}, and the chain rule gives
\begin{equation}
    \bm{f}_i^{T}=-\frac{\partial(A_TW_T)}{\partial\bm{x}_i}=-A_T\,\bm{P}_T\,\nabla N_i,
    \qquad \bm{P}_T=\bm{F}_T\bm{S}_T,
    \label{eq:fs}
\end{equation}
for each node $i$ of $T$. Here, $\bm{S}_T=\partial W_T/\partial\bm{\varepsilon}_T=\{Eh/(1-\nu^2)\}[(1-\nu)\bm{\varepsilon}_T+\nu(\mathrm{tr}\,\bm{\varepsilon}_T)\bm{I}]$ is the membrane force measured on the reference triangle (the second Piola--Kirchhoff stress resultant), $\bm{P}_T$ is the same force expressed in the current configuration (the first Piola--Kirchhoff stress resultant), and $\nabla N_i$ is the gradient of the linear shape function of node $i$ on the reference triangle, $\left(\nabla N_{i_1}\,\middle|\,\nabla N_{i_2}\right)=\bm{D}_{\rm m}^{-{\rm T}}$ and $\nabla N_{i_0}=-\nabla N_{i_1}-\nabla N_{i_2}$, so that the three forces sum to zero.

Given that the tensors $\bm{C}_{T,e}$ connecting curvature and dihedral angle in Eq.~\eqref{eq:Db} are constant, the bending energy depends on the node positions only through the dihedral angles, and the chain rule gives
\begin{equation}
    \bm{F}_i^{\rm b}=-\sum_e\frac{\partial U_{\rm b}}{\partial\theta_e}\,\frac{\partial\theta_e}{\partial\bm{x}_i},
    \qquad
    \frac{\partial U_{\rm b}}{\partial\theta_e}=\sum_{T\ni e}D\,A_T\left[(1-\nu)\,\Delta\bm{b}_T:\bm{C}_{T,e}+\nu\,(\mathrm{tr}\,\Delta\bm{b}_T)(\mathrm{tr}\,\bm{C}_{T,e})\right],
    \label{eq:fb}
\end{equation}
where the second sum runs over the two triangles sharing $e$ and $\Delta\bm{b}_T:\bm{C}_{T,e}=(\Delta b_T)_{\alpha\beta}(C_{T,e})_{\alpha\beta}$. The bracket in Eq.~\eqref{eq:fb} is the bending moment of triangle $T$ contracted with $\bm{C}_{T,e}$, so that $\partial U_{\rm b}/\partial\theta_e$ is the moment conjugate to the dihedral angle at $e$.

The dihedral-angle gradients are obtained by differentiating the definition of $\theta_e$ with respect to the node positions~\cite{Grinspun2003}. Let $e$ be an interior edge from node $\bm{x}_0$ to $\bm{x}_1$, of length $\ell_e=|\bm{x}_1-\bm{x}_0|$, shared by triangles $A=(\bm{x}_0,\bm{x}_1,\bm{x}_2)$ and $B=(\bm{x}_1,\bm{x}_0,\bm{x}_3)$ with unit normals $\bm{n}_A$ and $\bm{n}_B$ and heights $h_A=2\mathcal{A}_A/\ell_e$ and $h_B=2\mathcal{A}_B/\ell_e$ over the edge, where $\mathcal{A}_A$ and $\mathcal{A}_B$ are the current areas. Then
\begin{align}
    \frac{\partial\theta_e}{\partial\bm{x}_2}&=-\frac{\bm{n}_A}{h_A}, &
    \frac{\partial\theta_e}{\partial\bm{x}_3}&=-\frac{\bm{n}_B}{h_B},\nonumber\\
    \frac{\partial\theta_e}{\partial\bm{x}_0}&=(1-\alpha_A)\frac{\bm{n}_A}{h_A}+(1-\alpha_B)\frac{\bm{n}_B}{h_B}, &
    \frac{\partial\theta_e}{\partial\bm{x}_1}&=\alpha_A\frac{\bm{n}_A}{h_A}+\alpha_B\frac{\bm{n}_B}{h_B},
    \label{eq:dtheta}
\end{align}
where $\alpha_A=\hat{\bm{e}}\cdot(\bm{x}_2-\bm{x}_0)/\ell_e$ and $\alpha_B=\hat{\bm{e}}\cdot(\bm{x}_3-\bm{x}_0)/\ell_e$ are the positions of the feet of the two heights along the edge, measured from $\bm{x}_0$ in units of $\ell_e$.

\begin{figure}[!h]
    \centering
    \includegraphics[width=\textwidth]{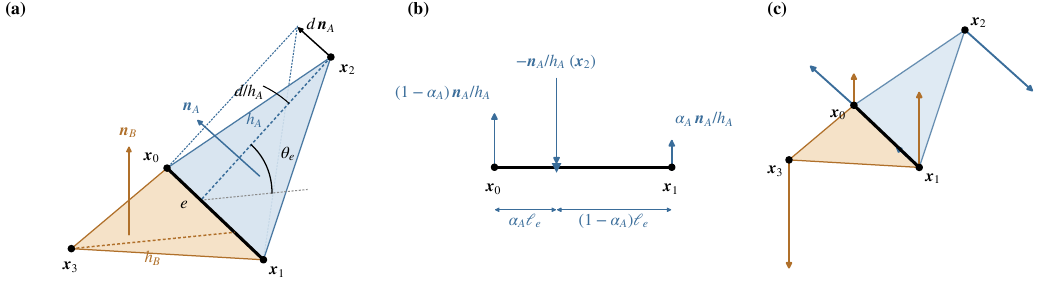}
    \caption{Geometry of the dihedral-angle gradients, Eq.~\eqref{eq:dtheta}. (a)~Triangles $A$ and $B$ sharing the edge $e$ from $\bm{x}_0$ to $\bm{x}_1$, with unit normals $\bm{n}_A$ and $\bm{n}_B$ and heights $h_A$ and $h_B$ over the edge; $\theta_e$ is the angle between the two triangles, i.e., between $\bm{n}_A$ and $\bm{n}_B$. Moving the vertex $\bm{x}_2$ by $d\bm{n}_A$ rotates $A$ about $e$ by $d/h_A$ (dashed). (b)~Lever rule for the contributions of $A$ along $\bm{n}_A$: the contribution $-\bm{n}_A/h_A$ of $\bm{x}_2$, acting at the foot of its height at $\alpha_A\ell_e$ from $\bm{x}_0$, is balanced by those of the edge nodes, so that rigid translations and rotations leave $\theta_e$ unchanged. The contributions of $B$ follow in the same way. (c)~Gradients $\partial\theta_e/\partial\bm{x}_i$ at the four nodes, with components along $\bm{n}_A$ (blue) and $\bm{n}_B$ (orange) drawn to scale.}    \label{fig:dihedral}
\end{figure}

These gradients have a simple geometric interpretation (Fig.~\ref{fig:dihedral}). Moving $\bm{x}_2$ by a small distance $d$ along $\bm{n}_A$, with the other nodes fixed, rotates triangle $A$ about the edge by $d/h_A$, since $h_A$ is the distance of $\bm{x}_2$ from the edge [Fig.~\ref{fig:dihedral}(a)], whereas displacements of $\bm{x}_2$ within the plane of $A$ leave $\theta_e$ unchanged; the same holds for $\bm{x}_3$. The gradients at the edge nodes ensure that rigid translations and rotations of the four nodes leave $\theta_e$ unchanged: the four gradients sum to zero, and the contribution of each opposite vertex, acting at the foot of its height, is shared between $\bm{x}_0$ and $\bm{x}_1$ by the lever rule [Fig.~\ref{fig:dihedral}(b)]. Figure~\ref{fig:dihedral}(c) shows the resulting gradients: those of the opposite vertices point against the normals of their own triangles, whereas those of the edge nodes combine the two normals with the weights of the lever rule.

Equation~\eqref{eq:eom} is integrated with a time step $\Delta t$ by updating the velocity with the damping treated by the trapezoidal rule, $\bm{v}_i\leftarrow[(1-\gamma\Delta t/2)\bm{v}_i+\Delta t\,\bm{F}_i/m]/(1+\gamma\Delta t/2)$, and then the position, $\bm{x}_i\leftarrow\bm{x}_i+\Delta t\,\bm{v}_i$.
In addition, we apply kinetic damping~\cite{Underwood1983}: when the total kinetic energy of the free nodes passes a maximum, i.e., when it decreases from one time step to the next, all their velocities are reset to zero and the motion restarts from rest. Since the kinetic energy peaks when the nodes pass the minimum of the potential energy along their current path, this removes energy efficiently and accelerates the convergence to equilibrium.
The velocities of the nodes on the boundaries are prescribed by the boundary conditions (Sec.~\ref{sec:specific}).

\subsection{Simulation Protocol}\label{sec:specific}

In the pinching simulations, the nodes at the edge points $(0,\pm R)$ of the mid-surface are displaced toward each other along $y$ by $\delta/2$ each, with their $x$ and $z$ coordinates fixed. The pinch $\delta$ is increased stepwise: the edge points are moved at a constant speed with smooth (cosine) ramps at both ends of each step, and are then held fixed while Eq.~\eqref{eq:eom} is integrated until the largest residual force on the free nodes and the kinetic energy fall below prescribed tolerances. We use the unit nodal mass, $m=1$, the time step $\Delta t=1/\omega_{\max}$, i.e., half the stability limit of the explicit scheme, where $\omega_{\max}$ is the largest eigenfrequency of the linearized stiffness at the reference configuration (estimated by power iteration), and the damping rate $\gamma=2\omega_1$, where $\omega_1$ is an estimate of the lowest bending frequency of a free plate of radius $R$. The tip displacement $u_{\rm tip}$ is the vertical displacement of the mid-surface at $(\pm R,0)$, averaged over the two tips. The simulations at finite pinch (Fig.~1 of the main text and Figs.~\ref{fig:S1} and \ref{fig:S2}) use $N_{\rm r}=40$ and increase $\delta$ by 0.5\,mm up to 5\,mm. The top surface in Fig.~1(c-i)--(e-i) of the main text is obtained by offsetting the mid-surface by $h/2$. The linear response in Figs.~2 and 3 of the main text is extracted from the first loading steps. For Fig.~2(a) and the elliptic planforms in Fig.~3(c), we use $N_{\rm r}=40$ and two small steps, $\delta=0.05$ and $0.1$\,mm, and take the linear coefficient $a$ of the fit $u_{\rm tip}=a\delta+b\delta^2$. For the parameter sweeps in Figs.~2(b) and 3(b) and for the circular planform in Fig.~3(c), we use $N_{\rm r}=20$ and take $u_{\rm tip}/\delta$ at the first step, $\delta=0.1$ or $0.5$\,mm; doubling $N_{\rm r}$ changes $u_{\rm tip}/\delta$ only by a few percent at most.

\subsection{Comparison with solid finite element method}\label{sec:validation}

We validate the discrete elastic shell against a three-dimensional finite-element model of the solid shell and against the experiments at the experimental thickness. In the solid model, the shell of thickness $h$ is discretized by quadratic (ten-node) tetrahedral elements with one element layer across the thickness, and a neo-Hookean material with the same Poisson's ratio is used. By symmetry, only a quarter of the shell is computed. The node at the top-surface corner of each edge point $(0,\pm R)$ is displaced toward the other by $\delta/2$, and the equilibrium is obtained by dynamic relaxation with lumped masses~\cite{Hinton1976, Underwood1983}.

\begin{figure}[!h]
    \centering
    \includegraphics[width=3.4in]{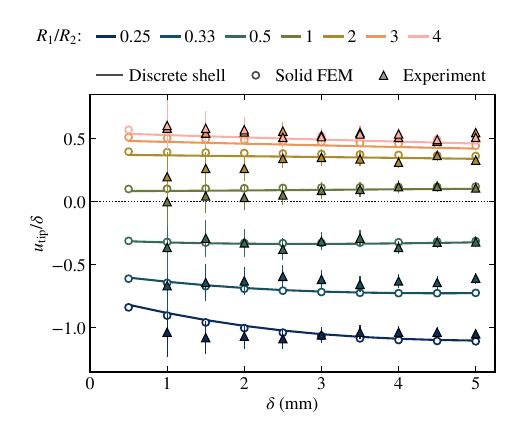}
    \caption{Validation of the discrete elastic shell at the experimental thickness, $h=0.6$\,mm ($R=R_2=20$\,mm). The normalized tip displacement $u_{\rm tip}/\delta$ is plotted against the pinch $\delta$ for the seven shapes used in the experiments, $R_1/R_2=0.25$--$4$ (colors). Lines: discrete elastic shell. Open circles: three-dimensional finite-element model of the solid shell. Filled triangles: experiments (error bars: resolution of the photographs).}
    \label{fig:S1}
\end{figure}

Figure~\ref{fig:S1} compares the normalized tip displacement $u_{\rm tip}/\delta$ for the seven shapes used in the experiments. The discrete shell and the solid model agree closely for all shapes and over the whole range of $\delta$: both reproduce the sign of the tip displacement and its weak dependence on $\delta$. The small differences between the two models, most visible for $R_1>R_2$, where the solid model gives a slightly larger rise, are smaller than the experimental uncertainty. The experiments follow the same trends, and the deviations that remain for some shapes are shared by both models; they are therefore not specific to the discrete-shell description.

\subsection{Calculation of rigidity landscape}\label{sec:rigidity}

For the rigidity landscape (Fig.~4 of the main text), we compute the linear response of the shell to a point force normal to its reference surface. As the reference mesh is stress-free, the linear response is governed by the stiffness matrix at the reference configuration,
\begin{equation}
    K_{ia,jb}=-\left.\frac{\partial F_{ia}}{\partial x_{jb}}\right|_{\bm{x}=\bm{X}},
    \label{eq:Kmat}
\end{equation}
where $F_{ia}$ is the $a$-th component ($a=x,y,z$) of the elastic force $\bm{F}_i^{\rm s}+\bm{F}_i^{\rm b}$ on node $i$ (Sec.~\ref{sec:detail}) and $x_{jb}$ is the $b$-th component of $\bm{x}_j$. We evaluate Eq.~\eqref{eq:Kmat} by central finite differences of the nodal forces, displacing each coordinate by $\pm10^{-6}$ times the mean edge length. As a displacement of a node changes the forces only on the nodes within two edges of it, nodes more than four edges apart are displaced simultaneously without their responses overlapping, so that the whole matrix is obtained from a few hundred force evaluations instead of six per node. The resulting matrix $\bm{K}$ is sparse and, as the Hessian of the elastic energy, symmetric up to round-off; we use its symmetric part, $(\bm{K}+\bm{K}^{\rm T})/2$.

The clamped arc, $3\pi/4\leq\varphi\leq5\pi/4$ with $\varphi$ the polar angle of the planform, is modeled by fixing all displacement components of the nodes on the two outermost circles of the mesh within the arc, which also fixes the slope at the edge; the rest of the edge is free. Fixing the clamped nodes removes the rigid-body motions, and the stiffness matrix restricted to the remaining (free) degrees of freedom, $\bm{K}_{\rm f}$, is positive definite. A unit force along $z$ at node $p$, represented by the unit vector $\bm{e}_p$ of the $z$ component of node $p$, produces the displacements $\bm{u}=\bm{K}_{\rm f}^{-1}\bm{e}_p$ of all free nodes, and the local stiffness is the inverse of the resulting deflection at $p$,
\begin{equation}
    k(p)=\frac{1}{u_z(p)}=\frac{1}{\bm{e}_p^{\rm T}\bm{K}_{\rm f}^{-1}\bm{e}_p}.
    \label{eq:kp}
\end{equation}
Note that $k(p)$ differs from the diagonal element of $\bm{K}_{\rm f}$, which would correspond to displacing node $p$ with all the other nodes held fixed.
Instead of forming the dense matrix $\bm{K}_{\rm f}^{-1}$, we factorize $\bm{K}_{\rm f}$ once into sparse triangular factors, $\bm{K}_{\rm f}=\bm{L}\bm{U}$ (LU decomposition), so that $u_z(p)$ at each probe point follows from one forward and one backward substitution, $\bm{L}\bm{y}=\bm{e}_p$ and $\bm{U}\bm{u}=\bm{y}$. Since forces can be applied only at nodes, each probe point is replaced by the nearest free node. Because the node spacing, $R/N_{\rm r}$, is smaller than the probe-point spacing, this shifts each probe point by less than one node spacing. The calculated stiffness $k$ is normalized as
\begin{equation}
    \frac{kR^2}{D\lambda}=12(1-\nu^2)\,\frac{kR_2}{Eh^2},
    \label{eq:knorm}
\end{equation}
so that values of order unity correspond to the stretching-dominated stiffness, $k\sim Eh^2/R_2$, discussed in the main text.

The geometry employed for this analysis is chosen to be that of the experiments, $R=R_2=20$\,mm with $R_2/R_1=-1$ (dome), 0 (cylinder), and 1 (saddle), and $\lambda=R^2/(R_2h)$ is varied through the thickness, $h=2.4$--$0.0375$\,mm ($\lambda=8.3$--$533$). The maps in Fig.~4(a) are evaluated at $\lambda=267$ on a square grid of spacing $0.04R$, using the mirror symmetry about $y=0$, with $N_{\rm r}=80$. The profiles in Fig.~4(b-i) are taken along two lines perpendicular to the lit--shadow boundary of the saddle, i.e., the asymptotic line from the upper end of the clamped arc to the center of the disk. The two lines cross the boundary at 35\% and 65\% of its length from the end of the arc ($\circ$ and $\square$, respectively in Fig.~4(a-iii)), and are sampled every $0.0125R$, with $s$ the signed distance from the boundary, positive into the shadow. The points in Fig.~4(b-ii) are $(x,y)/R=(-0.275,0)$, $(-0.525,0)$, and $(-0.6,0.2)$, inside the lit region.

\section{Stress profile}

\begin{figure}[!h]
    \centering
    \includegraphics[width=3.4in]{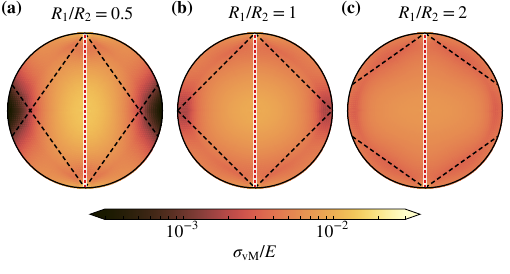}
    \caption{Surface stress of the pinched shell computed with the discrete elastic shell at the experimental thickness ($h=0.6$\,mm, $R=R_2=20$\,mm, $\delta=5$\,mm) for (a)~$R_1/R_2=0.5$, (b)~1, and (c)~2. The color shows the larger von Mises stress on the top and bottom surfaces, $\sigma_{\rm vM}$, normalized by Young's modulus $E$, on the undeformed planform. Dashed lines: asymptotic lines through the grip points, which bound the lit region. Red dotted line: grip line.}
    \label{fig:S2}
\end{figure}

Figure~\ref{fig:S2} shows the surface stress of pinched shells at the experimental thickness. For each triangle of the discrete shell, we evaluate the stresses on the top and bottom surfaces from the membrane force and the bending moment, $\bm{\sigma}^{\pm}=\bm{N}/h\pm6\bm{M}/h^2$, where $\bm{N}=\{Eh/(1-\nu^2)\}[(1-\nu)\bm{\varepsilon}_T+\nu(\mathrm{tr}\,\bm{\varepsilon}_T)\bm{I}]$ and $\bm{M}=D[(1-\nu)\Delta\bm{b}_T+\nu(\mathrm{tr}\,\Delta\bm{b}_T)\bm{I}]$ follow from the membrane strain and the curvature change of Eqs.~\eqref{eq:Us} and \eqref{eq:Ub}, and we plot the larger of the two von Mises stresses. The stress fills the lit region, takes intermediate values in the penumbra, and is much smaller in the shadow, which appears near the tips for $R_1<R_2$ [Fig.~\ref{fig:S2}(a)]. The largest stresses occur at the grip points and at the center. In all regions, the stress is dominated by bending rather than stretching.

This pattern follows from the isometric picture in the main text. The lit region inherits, along the asymptotic lines, the curvature of the bulge of the grip line ($F''\neq0$), which induces bending stresses. In the shadow, the function $F$ selected by bending decreases almost linearly beyond the asymptotic lines through the grip points, so that $F''\approx0$: the shadow tilts nearly rigidly, with little change in curvature and hence little stress. The penumbra carries one curved and one straight contribution and takes intermediate values. Thus the stress, as well as the displacement, follows the light-and-shadow rule.

\end{document}